\documentclass[aps, reprint, pra, superscriptaddress]{revtex4-2}

\usepackage{lineno}
\usepackage{microtype}
\usepackage[]{siunitx}
\usepackage{physics}
\usepackage{graphicx}
\usepackage{subcaption}
\usepackage{amsmath}
\usepackage{csquotes}
\providecommand\JournalTitle[1]{#1}

\usepackage{scalerel}
\newcommand*{\para}{\stretchrel*{\parallel}{\perp}}
\newcommand{\lng}{\protect{\para}}
\newcommand{\azm}{\protect{\perp}}

\begin{document}

\title{Diffusion of light in structurally anisotropic media with uniaxial symmetry}

\author{Ernesto Pini}
\email{pinie@lens.unifi.it}
\affiliation{Department of Physics and Astronomy, Universit\`{a} di Firenze, Sesto Fiorentino, Italy\\}
\affiliation{European Laboratory for Non-linear Spectroscopy (LENS), Sesto Fiorentino, Italy\\}
\author{Fabrizio Martelli}
\affiliation{Department of Physics and Astronomy, Universit\`{a} di Firenze, Sesto Fiorentino, Italy\\}
\author{Alexander Gatto}
\affiliation{Sony Europe B.V., Stuttgart Technology Center, Stuttgart, Germany\\}
\author{Henrik Sch\"afer}
\affiliation{Sony Europe B.V., Stuttgart Technology Center, Stuttgart, Germany\\}
\author{Diederik S. Wiersma}
\affiliation{Department of Physics and Astronomy, Universit\`{a} di Firenze, Sesto Fiorentino, Italy\\}
\affiliation{European Laboratory for Non-linear Spectroscopy (LENS), Sesto Fiorentino, Italy\\}
\affiliation{Istituto Nazionale di Ricerca Metrologica (INRiM), Turin, Italy\\}
\author{Lorenzo Pattelli}
\email{l.pattelli@inrim.it}
\affiliation{European Laboratory for Non-linear Spectroscopy (LENS), Sesto Fiorentino, Italy\\}
\affiliation{Istituto Nazionale di Ricerca Metrologica (INRiM), Turin, Italy\\}
\affiliation{National Research Council -- National Institute of Optics (CNR-INO), Sesto Fiorentino, Italy\\}

\date{\today}

\begin{abstract}
Anisotropic light transport is extremely common among scattering materials, yet a comprehensive picture of how macroscopic diffusion is determined by microscopic tensor scattering coefficients is not fully established yet. In this work, we present a theoretical and experimental study of diffusion in structurally anisotropic media with uniaxially symmetric scattering coefficients. Exact analytical relations are derived in the case of index-matched turbid media, unveiling the general relation between microscopic scattering coefficients and the resulting macroscopic diffusion tensor along different directions. Excellent agreement is found against anisotropic Monte Carlo simulations up to high degrees of anisotropy, in contrast with previously proposed approaches. The obtained solutions are used to analyze experimental measurements of anisotropic light transport in polystyrene foam samples under different degrees of uniaxial compression, providing a practical example of their applicability.
\end{abstract}

\maketitle

\section{Introduction}

Anisotropic light transport is a ubiquitous feature of many scattering media, ranging from biological materials as wood \cite{kienle2008light}, tendons \cite{kienle2007determination, simon2014time, kienle2004anisotropy}, teeth \cite{kienle2002light}, skin \cite{nickell2000anisotropy} and bone \cite{sviridov2005intensity}, to several types of plastics \cite{johnson2008full} and isotropic materials under mechanical deformations \cite{johnson2009optical}. Despite their relevance for many applications, however, anisotropic light transport properties are often overlooked or disregarded altogether, due to our incomplete understanding of how they arise from the structural features of a material.

In principle, for large scattering materials in the diffusive regime, a general Anisotropic Diffusion Equation (ADE) can be easily cast by replacing the scalar diffusion coefficient $D$ with a diffusion tensor $\vb*{D}$ containing the diffusive rates along different directions. In contrast with the isotropic case, however, the exact relation between the anisotropic diffusive coefficients and their corresponding microscopic scattering properties remains unknown, as well as the expression for the effective boundary conditions required to describe diffusion in finite media.

Despite the significance of structurally anisotropic materials and the need for accurate physical models to describe their light transport regime, a surprisingly small number of works exist which attempt to address this issue, with inconclusive results.
Up to date, the agreement between solutions to the radiative transfer equation and anisotropic diffusion has been mostly qualitative\cite{heino2003anisotropic}, with large discrepancies between Monte Carlo (MC) simulations and ADE which seemed to imply a fundamental irreducible incompatibility between anisotropic disorder and the core assumptions of diffusion theory \cite{kienle2007anisotropic}.
Attempts have been made to reconcile the two approaches, by suggesting a different relationship between scattering cross-sections and the components of the diffusion tensor \cite{johnson2009optical}, which however did not resolve the discrepancies with numerical Monte Carlo simulations \cite{kienle2013light}. In 2014, Alerstam showed that the observed deviations were due to the use of oversimplified assumptions on the interdependence of both the diffusion rates and the extrapolated boundary condition on the microscopic scattering coefficients \cite{alerstam2014anisotropic}, proposing a numerical recipe to compute all relevant transport quantities starting from a single anisotropic trajectory. When using the so obtained coefficients, excellent agreement was finally recovered between ADE and a general anisotropic Monte Carlo framework, albeit only up to a moderate degree of anisotropy. After this important milestone, only sporadic attempts were made towards the development of an analytical framework to connect microscopic and macroscopic transport parameters, without success \cite{han2020transport}.

In this work, we lay the foundation for an analytical description of anisotropic diffusion parameters based on microscopic scattering coefficients for materials characterized by uniaxial structural anisotropy. Closed-form relationships are derived in case of index-matched boundaries and no scattering asymmetry, which are found in excellent agreement with anisotropic Monte Carlo simulations up to very high degrees of anisotropy. Results are presented in the context of radiative transport, however, their derivation is general and could be seamlessly applied to other fields such as neutron transport in pebbled bed reactor cores \cite{vasques2014non} or particles in fluids \cite{tran2016simulation}, to name a few.

The paper is organized as follows: a theoretical background to anisotropic transport theory is provided in Section \ref{sec:ADE}, introducing the parameters of interest and the problem at hand. Section \ref{sec:AnSol} presents the derivation of analytical solutions in the case of an index-matched medium with uniaxially symmetric scattering tensor, allowing to evaluate the diffusive tensor elements and the extrapolated length directly from the scattering tensor coefficients. These solutions are validated against an open-source Monte Carlo implementation of anisotropic light transport in Section \ref{sec:MC}, and compared to time-resolved experimental measurements of diffuse reflectance from insulating foam samples under different degrees of controlled mechanical compression in Section \ref{sec:Exp}. Finally, the results are commented in Section \ref{sec:Con}.

\section{Anisotropic Diffusion Equation} \label{sec:ADE}
The three-dimensional time-dependent Anisotropic Diffusion Equation (ADE) for a scattering and absorbing medium can be written as:
\begin{equation}
	\left( \frac{1}{v} \pdv{t} - \div{\vb*{D}} \grad + \mu_\text{a} \right) \Phi(\vb*{r}, t) = S(\vb*{r}, t),
\end{equation}
where $\Phi(\vb*{r}, t)$
is the fluence rate at a position $\vb*{r} = (x, y, z)$ and time $t$, $\mu_\text{a}$ is the absorption coefficient, $S(\vb*{r}, t)$ is the source term, $v = c/n$ is the speed of light in the medium and $\vb*{D}$
is the anisotropic diffusion tensor
\begin{equation}
	\vb*{D} = \begin{pmatrix}
		D_x & 0 & 0 \\
		0 & D_y & 0 \\
		0 & 0 & D_z
	\end{pmatrix},
\end{equation}
having units of length.

At the microscopic level, we express the scattering anisotropy via a direction-dependent scattering coefficient $\mu_\text{s}(\vu{s})$
, which may exhibit a complex dependence on the unit vector $\vu{s}$ along the incoming direction of the energy packet before a scattering event.
On the other hand, in the following we will assume that the direction dependence of the scattering coefficient can be expressed using three scalar components of a scattering coefficient tensor $\vb*{\mu}_\text{s}$,
\begin{equation}
	\vb*{\mu}_\text{s} = \begin{pmatrix}
		\mu_{\text{s}, x} & 0 & 0 \\
		0 & \mu_{\text{s}, y} & 0 \\
		0 & 0 & \mu_{\text{s}, z}
	\end{pmatrix},
\end{equation}
through the relation $\mu_\text{s}(\vu{s}) = \vu{s} \vb*{\mu}_\text{s} \vu{s}^\text{T}$.
Similar tensors could be defined for the refractive index $n$ or the single-scattering asymmetry factor $g$. In fact, birefringence and direction-dependent scattering asymmetry may often be associated to the presence of structural anisotropy in tissues and fibrous materials.
All these tensors can be considered diagonal (allowing us to use a single subscript index for each component, for brevity), provided that a suitable set of absolute anisotropy axes can be defined in the laboratory reference frame, which in our case we will consider to be aligned with the Cartesian axes.
Even in this simplified scenario where each parameter can be defined by just three scalar components, however, the number of free parameters may still grow too large compared to the available experimental observables, making their full retrieval impractical.

In this work, for the sake of simplicity, we will assume that the transport anisotropy information can be condensed in the elements of the scattering coefficient tensor, assuming both $n$ and $g$ as direction-independent scalars, and further taking $g = 0$ to focus on the role of the scattering coefficient alone in shaping the anisotropic transport behavior.

Similarly, in analogy to the isotropic case, the scattering mean free paths $\ell_i$ and transport mean free paths $\ell_i^*$ along $i = \{x, y, z\}$ will be indicated as:
\begin{equation} \label{eq:ltdef}
	\mu_{\text{s}, i} = \frac{1}{\ell_i}, \qquad D_i = \frac{1}{3} \ell_i^* ,
\end{equation}
where it should be stressed that, despite setting $g = 0$, in general $\ell_i \neq \ell_i^*$ due to the presence of scattering anisotropy.
As a matter of fact, in contrast to the isotropic case, no separable ``similarity relation'' is expected to hold between the scattering and transport mean free paths, which will be mutually connected via a new adimensional anisotropy tensor $\vb*{K}$
\begin{equation}
	\vb*{K} = \begin{pmatrix}
		K_x & 0 & 0 \\
		0 & K_y & 0 \\
		0 & 0 & K_z
	\end{pmatrix},
\end{equation}
with elements defined via the relation:
\begin{equation} \label{eq:Kdef}
	\ell^*_i = K_i \ell_i .
\end{equation}
In other words, the tensor $\vb*{K}$ represents the transformation connecting the microscopic scattering properties to the macroscopic diffusion rates:
\begin{equation} \label{eq:tensorsimilarity}
	\vb*{D} = \frac{1}{3} \vb*{K} \vb*{\mu}_\text{s}^{-1}.
\end{equation}
In the isotropic case $\vb*{K}$ becomes the $3\times3$ identity matrix.

The diagonal elements of the diffusion tensor are the coefficients appearing in the anisotropic diffusion equation.
In this context, a commonly studied configuration is that of an infinitely extended slab with thickness $L$ illuminated by a pencil beam pulse along the perpendicular $z$ axis.
In the diffusive regime, a pencil beam source can be more conveniently modeled as an isotropic point source placed at the depth $z_0 = \ell_z$ from the surface $S(\vb*{r}, t) = \delta(x) \delta(y) \delta(z - z_0) \delta(t)$ (note that in the anisotropic case $z_0 = \ell_z \neq \ell_z^*$, even when assuming $g = 0$).
The solution of the time-dependent ADE can be written as:

\begin{widetext}
\begin{equation} \label{eq:Urt}
	\Phi(\vb*{r}, t) = \frac{v \exp(-\mu_\text{a}vt) \exp\left(-\frac{x^2}{4 D_x vt}\right) \exp\left(-\frac{y^2}{4 D_y vt}\right)}{(4 \pi vt)^{3/2} \sqrt{D_x D_y D_z}}
		\left[\sum_{m=-\infty}^{+\infty} \exp\left(-\frac{(z-z_{+, m})^2}{4 D_z vt}\right) - \exp\left(-\frac{(z-z_{-, m})^2}{4 D_z vt}\right)\right],
\end{equation}
\end{widetext}
with
\begin{align*}
	z_{+, m} &= 2 m \left( L+2 z_\text{e} \right) + z_0, \\
	z_{-, m} &= 2 m \left( L+2 z_\text{e} \right) - 2 z_\text{e} - z_0,
\end{align*}

and $z_\text{e}$ as the extrapolated length.
In order to solve the ADE as a function of the microscopic scattering coefficient elements $\mu_i$, closed-form expressions for the diffusive rates $D_i$ and extrapolated boundary length $z_\text{e}$ must be therefore derived.

\section{Analytical solutions for uniaxially symmetric anisotropy} \label{sec:AnSol}
A theoretical framework for the definition of a generalized relationship between scattering coefficients and diffusion rates has been presented recently, covering both non-classical propagation regimes and angle-dependent step length distributions \cite{vasques2014non}. This approach provides integral expressions which can be used to describe anisotropic transport \cite{martelli2022light}, even though explicit solutions to these expressions have not been reported nor validated against Monte Carlo simulations.
In the following, we present analytical solutions which are derived in the simplified case of media with uniaxially symmetric scattering coefficient tensor, with isotropic phase function ($g = 0$) and index-matched boundary conditions with the environment.
Under these conditions, simple closed-form expressions can be derived, which shed light on the analytical interdependence between microscopic and macroscopic (i.e., experimentally measurable) quantities.

Without loss of generality, let us define $z$ and $\phi$ as the anisotropy symmetry axis and the angle of rotation around it, respectively.
Consequently, we make a distinction between longitudinal (parallel to the symmetry axis) and azimuthal (perpendicular to the symmetry axis) parameters, renaming  $\mu_{\text{s}, z} = \mu_{\text{s}, \lng}$, $\mu_{\text{s}, x} = \mu_{\text{s}, y} = \mu_{\text{s}, \azm}$, and $D_z = D_\lng$, $D_x = D_y = D_\azm$.
Given a generic unit vector associated to the walker direction of propagation $\vu{s} = (\sin\theta \cos\phi, \sin\theta \sin\phi, \cos\theta)$, with $\theta$ and $\phi$ as the polar and azimuthal angles of the absolute (laboratory) reference frame, we find as expected that the direction-dependent scattering coefficient $\mu_\text{s}(\vu{s})$ is independent of the azimuthal angle $\phi$:
\begin{multline}
	\mu_\text{s}(\theta) = \mu_{\text{s}, \azm} \sin^2 \theta + \mu_{\text{s}, \lng} \cos^2 \theta = \\
	= \mu_{\text{s}, \azm} (1 - \chi^2) + \mu_{\text{s}, \lng} \chi^2 = \mu_\text{s}(\chi),
\end{multline}
where $\chi = \cos\theta$.

A first relevant quantity that can be derived is the expectation value of the mean free path $\expval{\ell}$
\begin{equation} \label{eq:ldef}
	\expval{\ell} = \int_{4\pi} \xi(\vu{s}) \ell(\vu{s}) \dd{\vu{s}},
\end{equation}
where $\xi(\vu{s})$ represents the probability that a walker travels within a solid angle $\dd{\vu{s}}$ about $\vu{s}$ after a scattering event, and $\ell(\vu{s}) = 1/\mu_\text{s}(\chi)$ is the direction-dependent scattering mean free path along the same direction \cite{martelli2022light}.
For an isotropic phase function, we get that $\xi(\vu{s}) = 1/4\pi$, so that the only source of anisotropy is given by $\ell(\vu{s})$ and the integral can be simplified to:
\begin{equation} \label{eq:l}
	\expval{\ell} = \frac{1}{2} \int_{-1}^{1} \frac{1}{\mu_\text{s}(\chi)} \dd{\chi} = \frac{\tanh^{-1}\sqrt{\frac{\mu_{\text{s}, \azm} - \mu_{\text{s}, \lng}}{\mu_{\text{s}, \azm}}}}{\sqrt{\mu_{\text{s}, \azm}}\sqrt{\mu_{\text{s}, \azm} - \mu_{\text{s}, \lng}}}.
\end{equation}

The diffusion coefficients, on the other hand, are defined as \cite{martelli2022light}:
\begin{align}
	D_\azm &= \frac{1}{4 \expval{\ell}} \int_{-1}^{1} \frac{1 - \chi^2}{\mu_\text{s}^2(\chi)} \dd{\chi}, \\
	D_\lng &= \frac{1}{2 \expval{\ell}} \int_{-1}^{1} \frac{\chi^2}{\mu_\text{s}^2(\chi)} \dd{\chi},
\end{align}
which also admit closed-form solutions:
\begin{widetext}
\begin{align}
	D_\azm &= \frac{1}{4} \left( \frac{2\mu_{\text{s}, \azm} - \mu_{\text{s}, \lng}}{\mu_{\text{s}, \azm}^2 - \mu_{\text{s}, \azm} \mu_{\text{s}, \lng}} - \frac{1}{\sqrt{\mu_{\text{s}, \azm}}\sqrt{\mu_{\text{s}, \azm} - \mu_{\text{s}, \lng}}\tanh^{-1} \sqrt{\frac{\mu_{\text{s}, \azm} - \mu_{\text{s}, \lng}}{\mu_{\text{s}, \azm}}}} \right) \label{eq:Dphi}, \\
	D_\lng &= \frac{1}{2} \left( \frac{1}{\mu_{\text{s}, \lng} - \mu_{\text{s}, \azm}} + \frac{\sqrt{\mu_{\text{s}, \azm}}}{\mu_{\text{s}, \lng}\sqrt{\mu_{\text{s}, \azm} - \mu_{\text{s}, \lng}}\tanh^{-1} \sqrt{\frac{\mu_{\text{s}, \azm} - \mu_{\text{s}, \lng}}{\mu_{\text{s}, \azm}}}} \right). \label{eq:Dperp}
\end{align}
\end{widetext}

From these analytical equations, it is evident that the macroscopic diffusion rate along a certain direction $D_i$ does not depend only on the scattering rates relative to the same direction $\mu_{\text{s}, i}$.
Based on equations \eqref{eq:tensorsimilarity} and \eqref{eq:l}, the coefficients can be rewritten in terms of the scattering mean free paths and the corresponding anisotropy tensor elements:
\begin{align}
	D_\azm  &= \frac{\ell_\azm}{4} \left[ 1 - \frac{\ell_\lng (\expval{\ell} - \ell_\azm)}{\expval{\ell} (\ell_\azm - \ell_\lng)} \right] = \frac{K_\azm \ell_\azm}{3}, \\
	D_\lng &= \frac{\ell_\lng}{2} \left[ \frac{\ell_\azm (\expval{\ell} - \ell_\lng)}{\expval{\ell} (\ell_\azm - \ell_\lng)} \right] = \frac{K_\lng \ell_\lng}{3},
\end{align}
with:
\begin{align}
	K_\azm  &= \frac{3}{4} \left[ 1 - \frac{\ell_\lng (\expval{\ell} - \ell_\azm)}{\expval{\ell} (\ell_\azm - \ell_\lng)} \right] \label{eq:Kphi}, \\
	K_\lng &= \frac{3}{2} \left[ \frac{\ell_\azm (\expval{\ell} - \ell_\lng)}{\expval{\ell} (\ell_\azm - \ell_\lng)} \right]. \label{eq:Kperp}
\end{align}
Notably, these equation reveal a conservation property for the trace of the anisotropy tensor:
\begin{equation} \label{eq:trK}
	\Tr(\vb*{K}) = K_\azm + K_\azm + K_\lng = 3,
\end{equation}
which holds independently of the degree of anisotropy between the longitudinal and perpendicular directions, and is verified to hold numerically also for full 3D anisotropy (i.e., when $\mu_{\text{s}, x} \neq \mu_{\text{s}, y} \neq \mu_{\text{s}, z}$).

The analytical solution for the extrapolated length $z_\text{e}$ is derived based on a general approach which accommodates for an anisotropic radiance \cite{alerstam2014anisotropic}.
Taking advantage of the absence of reflections at boundaries for an index-matched scattering medium, the only contributions determining the extrapolated length are given by the integrals $z_\text{e} = C/B$ \cite{alerstam2014anisotropic}:
\begin{align}
	B &= \int_0^{\pi/2} \dd{\theta} \int_0^{2\pi} \dd{\phi} P(\vu{s}) \sin\theta \cos\theta, \label{eqn:B} \\
	C &= \int_0^{\pi/2} \dd{\theta} \int_0^{2\pi} \dd{\phi} P(\vu{s}) \sin\theta \cos^2\theta \, \ell(\vu{s}), \label{eqn:C}
\end{align}
where $P(\vu{s})$ describes the angular distribution of the radiance, i.e., the probability density function that a random walker is traveling along a direction $\vu{s}$ after many scattering events.
In the general case, $P(\vu{s})$ depends both on the scattering rate tensor and the asymmetry factor $g$. In our case, we consider $g = 0$ so that to get $P(\vu{s}) = \ell(\vu{s}) / \expval{\ell}$, which further simplifies for uniaxial symmetry to $P(\chi) = 1/\mu_\text{s}(\chi) \expval{\ell}$.
We note that our equation \eqref{eqn:C} deviates slightly from that reported by Alerstam due to its dependence on $\ell(\vu{s})$ rather than $\ell^*(\vu{s})$ \cite{alerstam2014anisotropic}.
This change is motivated by the fact that we do not expect a simple similarity relation to hold in the case of anisotropic light transport. Moreover, irrespective of the value of the scattering asymmetry (which we have assumed here to be $g=0$), a general expression for the extrapolated length should depend on the microscopic scattering parameters rather than on the transport mean free path which, in the context of anisotropic propagation, is instead a direct expression of the macroscopic diffusion rates.
A validation of this correction against Monte Carlo simulations is discussed in Section \ref{sec:TRsteady}.

The previous integrals simplify to:
\begin{align}
	B &= \frac{2\pi v}{\expval{\ell}} \int_0^1 \frac{\chi}{\mu_\text{s}(\chi)} \dd{\chi}, \\
	C &= \frac{2\pi v}{\expval{\ell}} \int_0^1 \frac{\chi^2}{\mu^2_\text{s}(\chi)} \dd{\chi}.
\end{align}
to give an analytical expression for $z_\text{e}$
\begin{equation} \label{eq:ze}
	z_\text{e} = \frac{C}{B} = \frac{\frac{1}{\mu_{\text{s}, \lng}} - \frac{\tanh^{-1} \sqrt{(\mu_{\text{s}, \azm} - \mu_{\text{s}, \lng})/\mu_{\text{s}, \azm}}}{\sqrt{\mu_{\text{s}, \azm}} \sqrt{\mu_{\text{s}, \azm} - \mu_{\text{s}, \lng}}}}{\ln \frac{\mu_{\text{s}, \azm}}{\mu_{\text{s}, \lng}}} = \frac{\ell_\lng - \expval{\ell}}{\ln \frac{\ell_\lng}{\ell_\azm}}.
\end{equation}

Equations \eqref{eq:l}, \eqref{eq:Kphi}, \eqref{eq:Kperp} and \eqref{eq:ze} represent the main theoretical results introduced in this work.
As an important consistency check, it can be easily verified that all derived analytical equations converge to the familiar isotropic limit when $\mu_{\text{s}, \azm}$ and $\mu_{\text{s}, \lng}$ tend to the same value $\mu_\text{s} = 1/\ell$.

\section{Monte Carlo validation} \label{sec:MC}
We validate our analytic results using the open source Monte Carlo package PyXOpto \cite{naglic2021pyxopto}, which has been modified to handle anisotropic scattering and asymmetry factors.
The analytical ADE solution in the time and space domain is also provided as a set of MATLAB functions in the supplemental material.

\subsection{Transverse intensity spread}
Time-resolved reflected and transmitted intensity profiles for a slab illuminated along the $z$ axis can be obtained from the fluence rate in eq.\ \eqref{eq:Urt} using Fick's law \cite{martelli2022light, haskell1994boundary} by setting
\begin{align}
	R(x, y, t) &= D_z\frac{\partial}{\partial z}\Phi(x, y, z=0, t) \label{eq:Rrt}, \\
	T(x, y, t) &= -D_z\frac{\partial}{\partial z}\Phi(x, y, z=L, t) \label{eq:Trt} .
\end{align}
$R(x, y, t)$ and $T(x, y, t)$ are bi-variate Gaussian distributions:
\begin{align}
	R(x, y, t) &= R_0(t) \exp \left( -\frac{x^2}{4 D_x vt} \right) \exp \left( -\frac{y^2}{4 D_y vt} \right) \label{eq:bigauss}, \\
	T(x, y, t) &= T_0(t) \exp \left( -\frac{x^2}{4 D_x vt} \right) \exp \left( -\frac{y^2}{4 D_y vt} \right), \label{eq:bigauss2}
\end{align}
with mean square displacements (MSD) along $x$ and $y$
\begin{equation} \label{eq:MSDdef}
	w_x^2 (t) = 4 D_x vt, \qquad w_y^2(t) = 4 D_y vt
\end{equation}
and integrated time-domain amplitudes $R_0(t)$ and $T_0(t)$.

Accessing either $R(x, y, t)$ or $T(x, y, t)$ at different times allows to retrieve directly the diffusion rates (or, equivalently, the transport mean free paths $\ell_x^*$, $\ell_y^*$) along $x$ and $y$.
Moreover, the evolution of $w_x^2(t)$ and $w_y^2(t)$ does not depend on the amplitude of the profile, so that any effect that may modify its overall intensity (such as a homogeneous absorption coefficient) factors out exactly from the MSD.

\begin{figure}
	\centering
	\includegraphics{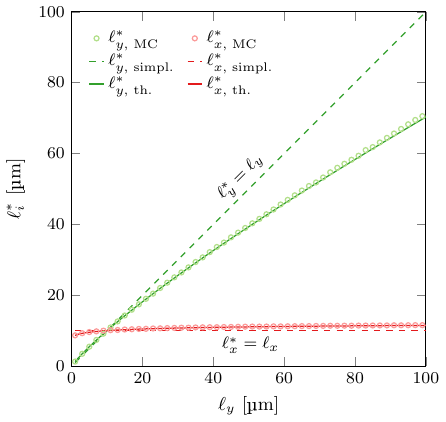}
	\caption{Mean free paths along $x$ and $y$ obtained with MC simulations and the analytical formulas as a function of $\ell_y$. Dashed lines represent the simplistic assumption $\ell^*_i = \ell_i$.}
	\label{fig:DxDy}
\end{figure}

We performed time- and space-resolved Monte Carlo simulations for an illustrative slab geometry of thickness $L=\SI[print-unity-mantissa=true]{1}{\milli\meter}$, setting a uniaxial scattering anisotropy along the $y$ axis ($\mu_{\text{s}, x} = \mu_{\text{s}, z} = \mu_{\text{s}, \azm}$, $\mu_{\text{s}, y} = \mu_{\text{s}, \lng}$).
We set $\ell_x = \ell_z = \SI{10}{\micro\meter}$, and vary $\ell_y$ between \SIrange{10}{100}{\micro\meter} with \SI{10}{\micro\meter} steps.
The macroscopic diffusion coefficients are retrieved by performing a linear regression on the growth rate of the MSD after an initial transient.
The resulting values of $\ell_x^*$ and $\ell_y^*$ are shown in Figure \ref{fig:DxDy} together with the analytical predictions obtained from equations \eqref{eq:Dphi} and \eqref{eq:Dperp}.
The deviation between theory and MC across this parameter range is about \SI{0.8}{\percent} along the longitudinal axis and \SI{0.2}{\percent} in the azimuthal plane, which is representative of the approximations of diffusion theory.
This discrepancy grows up to a maximum of \SI{1.4}{\percent} for the largest value of $\ell_y = \SI{100}{\micro\meter}$, corresponding to an extreme anisotropy ratio of $\ell_\lng / \ell_\azm = 10$. This is also expected in the context of diffusion theory, since a longer scattering mean free path along one axis will also increase the average scattering mean free path \eqref{eq:l}, hence reducing the ``effective'' optical thickness of the slab.

\subsection{Time-resolved and steady-state profiles} \label{sec:TRsteady}
Steady-state and total time-domain transmittance can be derived by direct integration of equation \eqref{eq:Trt} as
\begin{equation} \label{eq:Txy}
	T(x, y) = \int_0^\infty T(x, y, t) \dd{t},
\end{equation}
and
\begin{equation} \label{eq:Tt}
	T(t) = \iint_{-\infty}^\infty T(x, y, t) \dd{x} \dd{y},
\end{equation}
with analogue expressions for the reflectance.

\begin{figure}
\centering
	\includegraphics{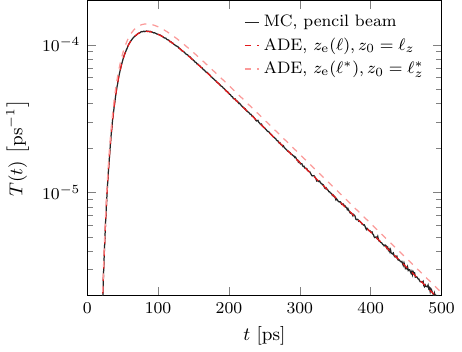}
	\caption{Total time-resolved transmittance through a \SI[print-unity-mantissa=true]{1}{\milli\meter}-thick slab with $\ell_x = \SI{30}{\micro\meter}$ and $\ell_y = \ell_z = \SI{10}{\micro\meter}$. Numerical results for pencil beam illumination are compared to the analytical solutions using different definitions for the extrapolated length $z_\text{e}$ and for the equivalent depth $z_0$ of the isotropic point source.}
	\label{fig:z0_ze}
\end{figure}

A first validation test regards the definition of the extrapolated length $z_\text{e}$ and the depth $z_0$ of the isotropic point source required to solve the ADE, which is such to match the case of a pencil beam illumination. In the current literature \cite{alerstam2014anisotropic}, it is suggested to compute $z_e$ as a function of  $\ell^*(\vu{s})$ and to use $z_0 = \ell_z^*$. However, when testing these assumptions against Monte Carlo simulations of anisotropic media with significant transport anisotropy, a much better agreement is obtained when defining $z_e$ as a function of $\ell(\vu{s})$ and setting $z_0 = \ell_z$ (Figure \ref{fig:z0_ze}).

\begin{figure*}
	\centering
	\includegraphics{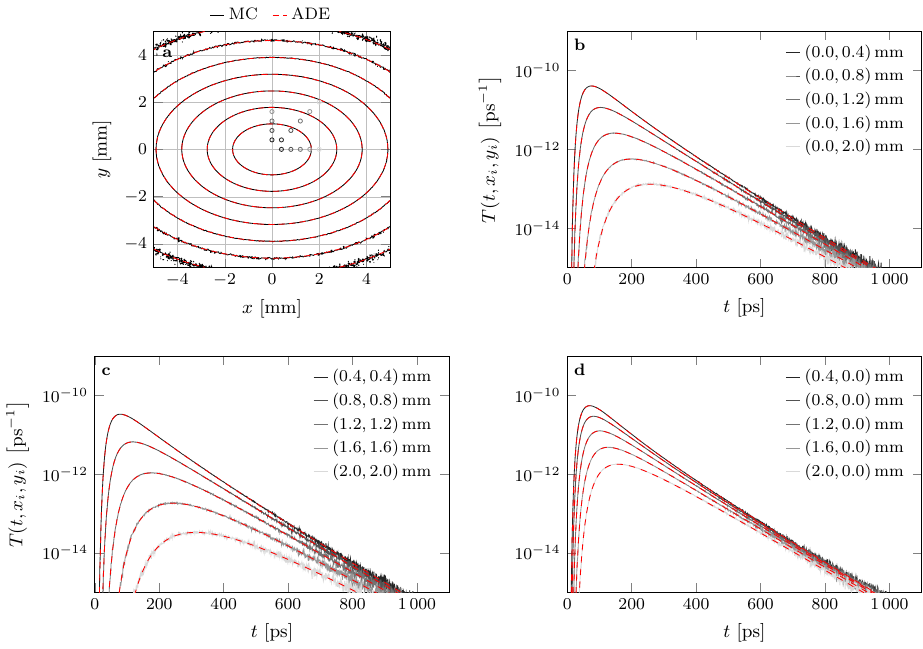}
	\caption{(a) Steady-state transmittance $T(x, y)$ from a \SI{1}{\milli\meter}-thick slab with $\ell_x = \SI{30}{\micro\meter}$ and $\ell_y = \ell_z = \SI{10}{\micro\meter}$, plotted as log-scale contour lines marking consecutive decades. (b, c, d) Time-resolved transmittance at different locations $(x_0, y_0)$ on the slab's output surface, identified by the markers in panel (a). The prediction from anisotropic theory (dashed) is compared against the results of a Monte Carlo simulation comprising \num{e13} trajectories (solid).}
	\label{fig:TSres}
\end{figure*}

A final set of MC simulations for the time- and space-resolved transmittance $T(x, y, t)$ was performed for slabs with $g = 0$, $n = 1$, $\mu_\text{a} = 0$ and different scattering coefficients. The results for one representative case are shown in Figure \ref{fig:TSres}, compared to the analytical predictions.
Excellent agreement is found at all times and positions, with average absolute discrepancies below \SI{1}{\percent} in the considered spatio-temporal range.

\section{Experimental measurements} \label{sec:Exp}
We finally test anisotropic light transport experimentally in samples of extruded polystyrene (XPS) foam under controlled degrees of uniaxial compression, and use the newly developed ADE to retrieve its microscopic scattering coefficients.
Due to its very high porosity, the effective refractive index of this material is close to that of air. Moreover, the absence of a well-defined boundary interface further suppresses the occurrence of internal reflections at boundaries, which conforms well with the assumptions under which our analytical results have been derived.

In pristine conditions, XPS foam is characterized by an isotropic closed-cell micro-structure and isotropic scattering properties.
Under uniaxial compression, however, the foam cells become increasingly oblate and eventually collapse, introducing different degrees of uniaxial scattering anisotropy \cite{krundaeva2016dynamic, vaitkus2013analysis, sadek2013finite}.

\begin{figure*}
	\centering
	\includegraphics{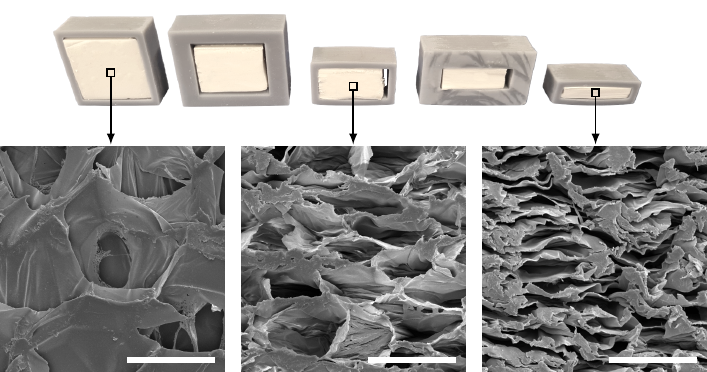}
	\caption{Experimental XPS samples under different compressive strain. The uncompressed height of each foam sample is $h_0 = \SI{29}{\milli\meter}$. A constant level of compression is maintained by placing the samples inside rigid frames. Scanning electron microscopy images of a cross-sectional slice of samples are shown in the bottom row for a few examples with increasing degree of compression. Scale bars correspond to \SI{100}{\micro\meter}.}
	\label{fig:samples}
\end{figure*}

\begin{figure*}
	\includegraphics[width=\linewidth]{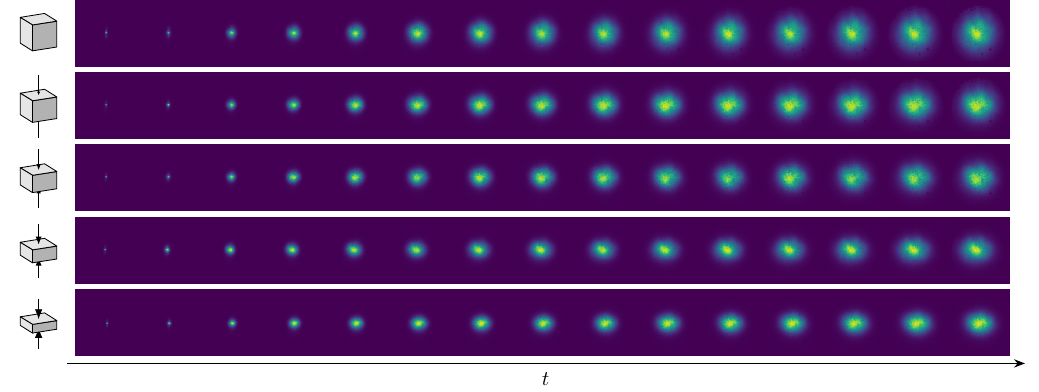}
	\caption{Time-resolved imaging measurements of the reflected intensity for the 5 samples of extruded polystyrene. The relative delay between consecutive frames is \SI[print-unity-mantissa=true]{1}{\pico\second}. Each frame covers an area of $5.35\times\SI{5.35}{\milli\meter\squared}$ and is normalized to its maximum value.}
	\label{fig:meas}
\end{figure*}

We realized different samples with increasing levels of compressive strain (defined as the ratio between the height contraction $\mathrm{\Delta} h$ and the original height $h_0$), spanning across the elastic-plastic deformation transition of XPS (Figure \ref{fig:samples})
Anisotropic light transport was studied by means of an optical gating imaging apparatus allowing to capture the evolution of intensity profiles with sub-\si{\pico\second} resolution \cite{pattelli2016spatio, pini2023breakdown}.
A collection of frames recorded for each sample at a probe wavelength of \SI{820}{\nano\meter} is shown in Figure \ref{fig:meas} for $t \leq \SI{15}{\pico\second}$. Within this time limited range, all samples can be effectively considered as semi-infinite.

Due to the uniaxial compression condition, each sample becomes characterized by a pair of azimuthal ($\ell_x = \ell_z$) and longitudinal ($\ell_y$) scattering mean free paths. We further assume $g = 0$ and no Fresnel reflections at boundaries, given the large pore sizes of the samples.
Nonetheless, an effective refractive index is calculated for each sample based on the increasing solid fraction under compression, ranging from $n = 1.014$ for the uncompressed case (solid fraction \SI{2.8}{\percent}) to $n = 1.043$ for the most compressed sample (solid fraction \SI{14}{\percent}). These values are used for the correct determination of the energy velocity $v$ inside the sample.

\begin{figure*}
	\centering
	\includegraphics{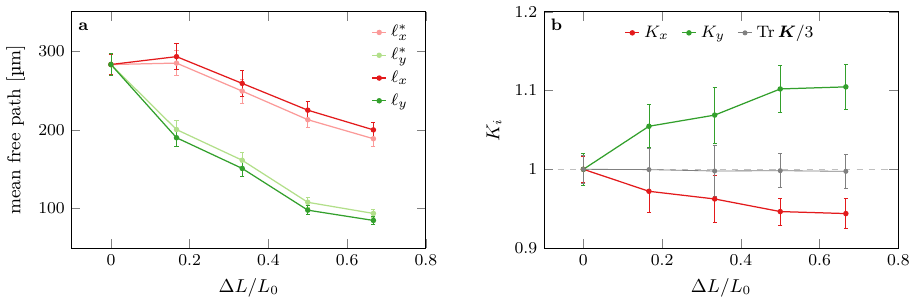}
	\caption{(a) Mean free path along $x$, $y$ and (b) anisotropy tensor components as a function of the compressive strain for the 5 samples of insulation foam.}
	\label{fig:ComprStrain}
\end{figure*}

The reflected intensity profiles were fitted at different times with bi-variate Gaussian distributions to retrieve their MSD evolution along the $y$ (vertical) and $x$ (horizontal) axes.
The resulting diffusive constants $D_x$, $D_y$, expressed through their corresponding transport mean free paths, can be then obtained via a linear regression of the MSD growth rates.
The microscopic scattering coefficients $\ell_x$ and $\ell_y$ along the two anisotropic directions can be finally retrieved using the analytical relations \eqref{eq:Dperp} and \eqref{eq:Dphi}. The results are shown in Figure \ref{fig:ComprStrain}a.

As intuitively expected, we find that $\ell_y$ is progressively reduced with increasing compression.
On the other hand, $\ell_x = \ell_z$ remains mostly constant for small degrees of compression, and then starts to decrease with increasing compressive strain (Fig.~\ref{fig:ComprStrain}a).
The overall variation is well captured by the anisotropy tensor elements introduced in equation \eqref{eq:Kdef}, which exhibit opposite trends as a function of compressive strain. As suggested by the relation $\Tr(\vb*{K})/3 = 1$, the increase of $K_y$ is compensated by a corresponding decrease of $K_x$ (Fig.~\ref{fig:ComprStrain}b).
This analysis allows to quantitatively estimate the magnitude of the systematic error introduced when inferring microscopic scattering parameters directly from macroscopic observables in structurally anisotropic media, which may directly affects several results previously reported in the literature \cite{burresi2014bright, cortese2015anisotropic, utel2019optimized, park2022surpassing}.

\section{Conclusions} \label{sec:Con}
In this work, we analyzed anisotropic light transport induced by a uniaxial tensor scattering coefficient, and derived analytic expressions for both the diffusion rates and the extrapolated boundary length.
In the simple case of isotropic scattering ($g = 0$) and index-matched boundary conditions, the resulting closed-form expression allow to solve directly the anisotropic diffusion equation both in the spatial and temporal domains.
The results have been validated against a general anisotropic Monte Carlo implementation, showing excellent agreement up to high degrees of anisotropy between the longitudinal and azimuthal directions.

The availability of an analytic solution to the anisotropic diffusion equation can streamline the analysis of experimental data, which we demonstrate in the case of a scattering medium subject to a controlled degree of anisotropic deformation.
Additionally, it provides a direct insight into the functional interdependence between the scattering properties along different axes, which may be extended to more general anisotropic configurations.
More importantly, our study revealed the presence of systematic discrepancies between the observed diffusive rates and the corresponding scattering coefficients, calling for a careful re-evaluation of previously reported scattering mean free path values of anisotropic media.
At the same time, it facilitates the application of a more correct diffusive model in several cases where structural anisotropy is evident and yet disregarded due to the lack of established numerical or theoretical approaches.

The analytical solution that we have presented is applicable only to the specific case of index-matched media with an isotropic phase function.
This represents a configuration of prominent relevance, especially considering the fact that in the diffusive regime we do not expect to be able to uniquely determine the (tensor) scattering asymmetry from the diffusive rates.
Indeed, experimental studies on anisotropic light transport typically consider only one tensor for the scattering coefficients, making our approach directly applicable to the analysis of a large array of previously reported results.
Additionally, as we have shown, there is potentially a large class of materials characterized by suppressed reflection at boundaries and uniaxial anisotropy, such as foams or aerogels which are permeated by their host medium, but also biological tissues which are immersed in an aqueous environment decreasing their index contrast with the surrounding media.

For more general anisotropic media, with different scattering and asymmetry coefficients along each axis, we anticipate that closed-form expressions may very likely be impractical to use or not exist altogether.
However, even in this case, our results confirm the validity of the adopted theoretical framework for the calculation of the relevant transport parameters, which may still be solved numerically for more complex configurations at a fraction of the computational cost required to run a full anisotropic Monte Carlo simulation, driving the further development of new solutions for the accurate description of direction-dependent diffusion processes.

\section*{Funding}
This work was partially funded by the European European Union's NextGenerationEU Programme with the I-PHOQS Research Infrastructure [IR0000016, ID D2B8D520, CUP B53C22001750006] ``Integrated infrastructure initiative in Photonic and Quantum Sciences''. E.P acknowledges financial support from Sony Europe B.V.. L.P. further acknowledges the CINECA award under the ISCRA initiative, for the availability of high performance computing resources and support (ISCRA-C ``ARTTESC''), and NVIDIA Corporation for the donation of the Titan X Pascal GPU.
F.M. acknowledges financial support by the European Union's - NextGenerationEU, National Recovery and Resilience Plan, MNESYS, PE0000006 (DN 572 1553 11.10.2022) and PRIN 2022, DIRS, grant 578 number: 2022EB4B7E.

\section*{Acknowledgments}
M. B\"{u}rmen and P. Nagli\v{c} are kindly acknowledged for fruitful discussion.

%\bibliography{apssamp}

\end{document}